\documentstyle[prb,aps,epsf,multicol]{revtex}
\newif\ifmynarrow \mynarrowfalse
\renewcommand{\narrowtext}{%
  \ifmynarrow\hspace*{\fill}\raisebox{-1ex}[0pt][0pt]{%
    \rule{0.3pt}{1ex}%
    \rule[1ex]{20.5pc}{0.3pt}}\fi
  \mynarrowtrue
  \vspace{-1.0ex}%
  \begin{multicols}{2}%
  \par\global\columnwidth20.5pc
  \global\hsize\columnwidth\global\linewidth\columnwidth
  \global\displaywidth\columnwidth}

\renewcommand{\widetext}{%
  \end{multicols}%
  \vspace{-2.5ex}%
  \noindent\raisebox{1ex}[0pt][0pt]{%
    \rule{20.5pc}{0.3pt}%
    \rule{0.3pt}{1ex}}%
  \par\global\columnwidth42.5pc
  \global\hsize\columnwidth\global\linewidth\columnwidth
  \global\displaywidth\columnwidth}

\begin{document}

\title{Rashba spin splitting in two-dimensional electron and hole systems}

\draft

\author{R. Winkler}
\address{Institut f\"ur Technische Physik III, Universit\"at
Erlangen-N\"urnberg, Staudtstr. 7, D-91058 Erlangen, Germany}

\date{February 1, 2000}
\maketitle
\begin{abstract}
  In two-dimensional (2D) hole systems the inversion asymmetry
  induced spin splitting differs remarkably from its familiar
  counterpart in the conduction band. While the so-called Rashba
  spin splitting of electron states increases linearly with in-plane
  wave vector $k_\|$ the spin splitting of heavy hole states can be
  of third order in $k_\|$ so that spin splitting becomes negligible
  in the limit of small 2D hole densities. We discuss consequences
  of this behavior in the context of recent arguments on the origin
  of the metal-insulator transition observed in 2D systems.
\end{abstract}

\narrowtext

At zero magnetic field $B$ spin splitting in quasi two-dimensional
(2D) semiconductor quantum wells (QW's) can be a consequence of the
bulk inversion asymmetry (BIA) of the underlying crystal (e.g.\ a
zinc blende structure) and of the structure inversion asymmetry
(SIA) of the confinement potential. This $B=0$ spin splitting
\cite{kit63,chiral} is the subject of considerable interest because
it concerns details of energy band structure that are important in
both fundamental research and electronic device applications (Refs.\ 
\onlinecite{pfe99,bych84,win93,eke85,win96,pud97,sko98,lom85,ros89,%
wol89,luo90,nit97,hei98,eis84,las85,dre92,jus92,lu98,pap99} and references
therein).

Here we want to focus on the SIA spin splitting which is usually the
dominant part of $B=0$ spin splitting in 2D systems. \cite{pfe99} To
lowest order in $k_\|$ SIA spin splitting in 2D electron systems is
given by the so-called Rashba model, \cite{bych84} which predicts a
spin splitting linear in $k_\|$.  For small in-plane wave vector
$k_\|$ this is in good agreement with more accurate numerical
computations. \cite{win93} For 2D hole systems, on the other hand,
the situation is more complicated because of the fourfold degeneracy
of the topmost valence band $\Gamma_8^v$, and so far only numerical
computations on hole spin splitting have been performed.
\cite{eke85,win96} In the present paper we will develop an
analytical model for the SIA spin splitting of 2D hole systems. We
will show that in contrast to the familiar Rashba model the spin
splitting of heavy hole (HH) states is basically proportional to
$k_\|^3$. This result was already implicitly contained in several
numerical computations. \cite{eke85,win96} But a clear analytical
framework was missing. We will discuss consequences of this behavior
in the context of recent arguments on the origin of the
metal-insulator transition observed in 2D systems.
\cite{pud97,sko98}

First we want to review the major properties of the Rashba model
\cite{bych84}
\begin{equation}
\label{rashbafull}
H_{6c}^{\rm SO} = \alpha \, {\bf k} \times {\bf E} \cdot
\bbox{\sigma} .
\end{equation}
In this equation $\bbox{\sigma} = (\sigma_x, \sigma_y, \sigma_z)$
denotes the Pauli spin matrices, $\alpha$ is a material-specific
prefactor, \cite{lom85,ros89} and ${\bf E}$ is an effective electric
field that results from the built-in or external potential $V$ as
well as from the position dependent valence band edge. For ${\bf E}
= (0,0,E_z)$ Eq.\ (\ref{rashbafull}) becomes (using explicit matrix
notation)
\begin{equation}
\label{rashba}
H_{6c}^{\rm SO} = \alpha \, E_z \left(
  \begin{array}{cc} 0 & k_- \\ k_+ & 0 \end{array}
  \right)
\end{equation}
with $k_\pm = k_x \pm i k_y$. By means of perturbation theory we
obtain for the spin splitting of the energy dispersion
\begin{equation}
\label{rashbaperturb}
{\cal E}_{6c}^{\rm SO} ({\bf k}_\|) =
\pm \langle \alpha E_z \rangle  k_\|
\end{equation}
where ${\bf k}_\| = (k_x, k_y, 0)$. Using this simple formula
several groups determined the prefactor $\langle \alpha E_z \rangle
$ by analyzing Shubnikov-de Haas (SdH) oscillations.
\cite{wol89,luo90,nit97,hei98}

\begin{figure}
\centerline{\epsfxsize=0.80\columnwidth\leavevmode
 \epsffile{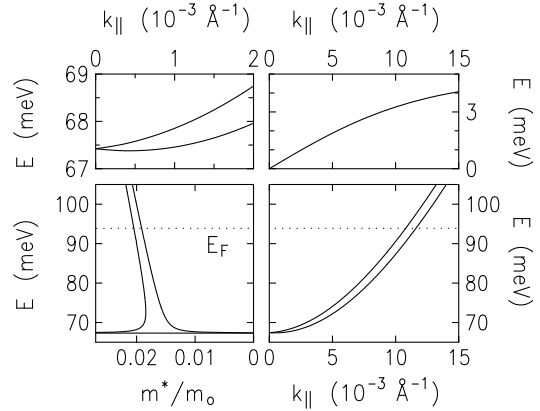}}\vspace{3mm}
\caption[]{\label{pic:insb} Self-consistently calculated subband
dispersion ${\cal E}_\pm (k_\|)$ (lower right), DOS
effective mass $m^\ast/m_0$ (lower left), spin splitting ${\cal E}_+
(k_\|) - {\cal E}_- (k_\|)$ (upper right) and subband dispersion
${\cal E}_\pm (k_\|)$ in the vicinity of $k_\|=0$ (upper left) for
an MOS inversion layer on InSb with $N_s = 2 \times 10^{11}$
cm$^{-2}$ and $|N_A - N_D| = 2\times 10^{16}$ cm$^{-2}$. The dotted
line indicates the Fermi energy $E_F$.}
\end{figure}

Equation (\ref{rashbaperturb}) predicts an SIA spin splitting which
is linear in $k_\|$. For small $k_\|$ Eq.\ (\ref{rashbaperturb})
thus becomes the dominant term in the energy dispersion ${\cal
E}_\pm ({\bf k_\|})$, i.e., SIA spin splitting of electron states is
most important for small 2D densities. In particular, we get a
divergent van Hove singularity of the density-of-states (DOS) at the
bottom of the subband \cite{win93} which is characteristic for a $k$
linear spin splitting. As an example, we show in Fig.\ 
\ref{pic:insb} the self-consistently calculated \cite{win93} subband
dispersion ${\cal E}_\pm (k_\|)$, DOS effective mass $m^\ast/m_0$,
and spin splitting ${\cal E}_+ (k_\|) - {\cal E}_- (k_\|)$ for an
MOS inversion layer on InSb. For small $k_\|$ the spin splitting
increases linearly as a function of $k_\|$, in agreement with Eq.\ 
(\ref{rashbaperturb}). Due to nonparabolicity the spin splitting for
larger $k_\|$ converges toward a constant. \cite{win93}

The spin splitting results in unequal populations $N_\pm$ of the two
branches ${\cal E}_\pm (k_\|)$. For a given total density $N_s = N_+
+ N_-$ and a subband dispersion ${\cal E}_\pm (k_\|) = \langle \mu
\rangle k_\|^2 \pm \langle \alpha E_z \rangle k_\|$ with $\mu =
\hbar^2/2m^\ast$ we obtain
\begin{equation}
\label{sdh_rashba}
N_\pm = \frac{1}{2} N_s \pm \frac{\langle \alpha E_z \rangle}{8\pi
\langle \mu \rangle^2} \sqrt{8\pi \langle \mu \rangle^2 N_s -
\langle \alpha E_z \rangle^2} .
\end{equation}
This equation can be directly compared with, e.g., the results of
SdH experiments. \cite{wol89,luo90,nit97,hei98}

The Rashba model (\ref{rashbafull}) can be derived by purely
group-theoretical means. The electron states in the lowest
conduction band are $s$ like (orbital angular momentum $l=0$). With
spin-orbit (SO) interaction we have total angular momentum $j=1/2$.
Both ${\bf k}$ and ${\bf E}$ are polar vectors and ${\bf k} \times
{\bf E}$ is an axial vector (transforming according to the
irreducible representation $\Gamma_4$ of $T_d$). \cite{ros89,trr79}
Likewise, the spin matrices $\sigma_x$, $\sigma_y$, and $\sigma_z$
form an axial vector $\bbox{\sigma}$. The dot product
(\ref{rashbafull}) of ${\bf k} \times {\bf E}$ and $\bbox{\sigma}$
therefore transforms according to the identity representation
$\Gamma_1$, in accordance with the theory of invariants of Bir and
Pikus. \cite{bir74} In the $\Gamma_6^c$ conduction band the scalar
triple product (\ref{rashbafull}) is the only term of first order in
${\bf k}$ and ${\bf E}$ that is compatible with the symmetry of the
band.

Now we want to compare the Rashba model (\ref{rashbafull}) with the
SIA spin splitting of hole states. The topmost valence band is $p$
like ($l=1$). With SO interaction we have $j=3/2$ for the HH/LH
states ($\Gamma_8^v$) and $j=1/2$ for the SO states ($\Gamma_7^v$).
For the $\Gamma_8^v$ valence band there are two sets of matrices
which transform like an axial vector, namely ${\bf J} =
(J_x,J_y,J_z)$ and $\bbox{\cal J} = (J_x^3,J_y^3,J_z^3)$ (Refs.\ 
\onlinecite{trr79,lut56}). Here $J_x$, $J_y$ and $J_z$ are the
angular momentum matrices for $j=3/2$. Thus we get \cite{cross}
\begin{equation}
\label{lutt_spin}
H_{8v}^{\rm SO} = \beta_1 \, {\bf k} \times {\bf E} \cdot {\bf J} +
\beta_2 \, {\bf k} \times {\bf E} \cdot \bbox{\cal J} .
\end{equation}
Similar to the Rashba model the first term has axial symmetry with
the symmetry axis being the direction of the electric field ${\bf
E}$. The second term is anisotropic, i.e., it depends on both the
crystallographic orientation of ${\bf E}$ and ${\bf k}$. Using ${\bf
k} \cdot {\bf p}$ theory we find that the prefactor $\beta_2$ is
always much smaller than $\beta_1$, i.e., the dominant term in Eq.\ 
(\ref{lutt_spin}) is the first term. This can be easily understood
by noting that the ${\bf k} \cdot {\bf p}$ coupling between
$\Gamma_8^v$ and $\Gamma_6^c$ is isotropic, so that it contributes
to $\beta_1$ but not to $\beta_2$. The prefactor $\beta_2$ stems
from ${\bf k} \cdot {\bf p}$ coupling to more remote bands such as
the $p$ antibonding conduction bands $\Gamma_8^c$ and $\Gamma_7^c$.

For ${\bf E} = (0,0,E_z)$ Eq.\ (\ref{lutt_spin}) becomes (using
explicit matrix notation with $j=3/2$ eigenstates in the order $j_z
= +3/2,+1/2,-1/2,-3/2$)
\widetext
\begin{equation}
\label{lutt_spin_x}
H_{8v}^{\rm SO} = \beta_1 \, E_z
\left(\begin {array}{cccc}
 0 & \frac{1}{2} \sqrt{3} \, k_- & 0 & 0 \\[0.5ex]
 \frac{1}{2} \sqrt{3} \, k_+ & 0 & k_- & 0 \\[0.5ex]
 0 & k_+ & 0 & \frac{1}{2} \sqrt{3} \, k_- \\[0.5ex]
 0 & 0 & \frac{1}{2} \sqrt{3} \, k_+ & 0 \end {array} \right)
+ \beta_2 \, E_z
\left(
\begin {array}{cccc}
 0 & {\frac{7}{8}} \sqrt {3} \, k_- & 0 & 3/4 \, k_+ \\[0.5ex]
 {\frac {7}{8}} \sqrt {3} \, k_+ & 0 & 5/2 \, k_- & 0 \\[0.5ex]
 0 & 5/2 \, k_+ & 0 & {\frac {7}{8}} \sqrt {3} \, k_- \\[0.5ex]
 3/4 \, k_- & 0 & {\frac {7}{8}} \sqrt {3} \, k_+ & 0
\end {array} \right) .
\end{equation}
\narrowtext\noindent
Here the first term couples the two LH states ($j_z = \pm 1/2$), and
it couples the HH states ($j_z = \pm 3/2$) with the LH states. But
there is no $k$ linear splitting of the HH states proportional to
$\beta_1$. The second matrix in Eq.\ (\ref{lutt_spin_x}) contains a
$k$ linear coupling of the HH states.

We want to emphasize that $H_{6c}^{\rm SO}$ and $H_{8v}^{\rm SO}$
are {\em effective} Hamiltonians for the spin splitting of electron
and hole subbands, which are implicitly contained in the full
multiband Hamiltonian for the subband problem \cite{win93,las85}
\begin{equation}
\label{kp_sub}
H = H_{{\bf k} \cdot {\bf p}} ({\bf k}_\|, k_z = -i \partial_z)
    + e E_z z \openone .
\end{equation}
Here $H_{{\bf k} \cdot {\bf p}}$ is a ${\bf k} \cdot {\bf p}$
Hamiltonian for the bulk band structure (i.e., $H_{{\bf k} \cdot
{\bf p}}$ does not contain $H_{6c}^{\rm SO}$ or $H_{8v}^{\rm SO}$)
and we have restricted ourselves to the lowest order term in a
Taylor expansion of the confining potential $V(z) = V_0 + eE_z z +
{\cal O} (z^2)$ which reflects the inversion asymmetry of $V(z)$.
The effective Hamiltonians (\ref{rashba}) and (\ref{lutt_spin_x})
stem from the combined effect of $H_{{\bf k} \cdot {\bf p}}$ and the
term $e E_z z$. For a systematic investigation of the importance of
the different terms in $H$ we have developed a novel, analytical
approach based on a perturbative diagonalization of $H$ using a
suitable set of trial functions and using L\"owdin partitioning.
\cite{bir74,loe51} Though we cannot expect accurate numerical
results from such an approach it is an instructive complement for
numerical methods, as we can clearly identify in the subband
dispersion ${\cal E} ({\bf k}_\|)$ the terms proportional to $E_z$
which are breaking the spin degeneracy. Neglecting in $H_{{\bf k}
\cdot {\bf p}}$ remote bands like $\Gamma_8^c$ and $\Gamma_7^c$ we
obtain for the SIA spin splitting of the HH states
\begin{mathletters}
\label{luttperturb}
\begin{equation}
\label{luttperturbHH}
{\cal E}_{\rm HH}^{\rm SO} (k_\|) \propto
\pm \langle \beta_1 E_z \rangle  k_\|^3 .
\end{equation}
In particular, we have no $k$ linear splitting (and $\beta_2 \equiv
0$) if we restrict ourselves to the Luttinger Hamiltonian
\cite{lut56} which includes $\Gamma_8^c$ and $\Gamma_7^c$ by means
of second order perturbation theory. \cite{eke85} Accurate numerical
computations \cite{win93} show that the dominant part of the $k$
linear splitting of the HH states is due to BIA. However, for
typical densities this $k$ linear splitting is rather small. For the
LH states we have
\begin{equation}
\label{luttperturbLH}
{\cal E}_{\rm LH}^{\rm SO} (k_\|) \propto
\pm \langle \beta_1 E_z \rangle  k_\| .
\end{equation}
\end{mathletters}%
Thus we have a qualitative difference between the spin splitting of
electron and LH states which is proportional to $k_\|$ and the
splitting of HH states which essentially is proportional to
$k_\|^3$. The former is most important in the low-density regime
whereas the latter becomes negligible for small densities. Note that
for 2D hole systems the first subband is HH like so that for low
densities the SIA spin splitting is given by Eq.\ 
(\ref{luttperturbHH}). In Eq.\ (\ref{luttperturb}) the lengthy
prefactors depend on the details of the geometry of the QW.
Moreover, we have omitted a weak dependence on the direction of
${\bf k}_\|$. But the order of the terms with respect to $k_\|$ is
independent of these details. It is crucial that, basically, we have
\begin{equation}
\label{so_bulk}
\alpha, \beta_1, \beta_2 \propto \Delta_0
\end{equation}
with $\Delta_0$ the SO gap between the bulk valence bands
$\Gamma_8^v$ and $\Gamma_7^v$, i.e., we have no SIA spin splitting
for $\Delta_0 = 0$. \cite{ros89} This can be most easily seen if we
express $H_{{\bf k} \cdot {\bf p}}$ in a basis of orbital angular
momentum eigenstates.

A more detailed analysis of our analytical model shows that both
$H_{6c}^{\rm SO}$ and $H_{8v}^{\rm SO}$ stem from a third order
perturbation theory for $k_\pm$, $k_z = -i \partial_z$, and $e E_z
z$. This seems to be a rather high order. Nevertheless, the
resulting terms are fairly large. \cite{gap} In agreement with
Refs.\ \onlinecite{pfe99,win93,las85} this is a simple argument to
resolve the old controversy based on an argument by Ando
\cite{afs82} that spin splitting in 2D systems ought to be
negligibly small because for bound states in first order we have
$\langle E_z \rangle = 0$. We note that the present ansatz for the
prefactors $\alpha$ and $\beta_1, \beta_2$ is quite different from
the ansatz in Ref.\ \onlinecite{lom85}. We obtain $H_{6c}^{\rm SO}$
and $H_{8v}^{\rm SO}$ by means of L\"owdin partitioning of the
Hamiltonian (\ref{kp_sub}) whereas in Ref.\ \onlinecite{lom85} the
authors explicitly introduced $H_{6c}^{\rm SO}$ into their model.
Moreover, we evaluate the matrix elements of $eE_z z$ with respect
to envelope functions for the bound states whereas in
Ref.~\onlinecite{lom85} the authors considered matrix elements of
$eE_z z$ with respect to bulk Bloch functions. The latter quantities
are problematic because they depend on the origin of the coordinate
frame.

\begin{figure}
\centerline{\epsfxsize=0.85\columnwidth\leavevmode
 \epsffile{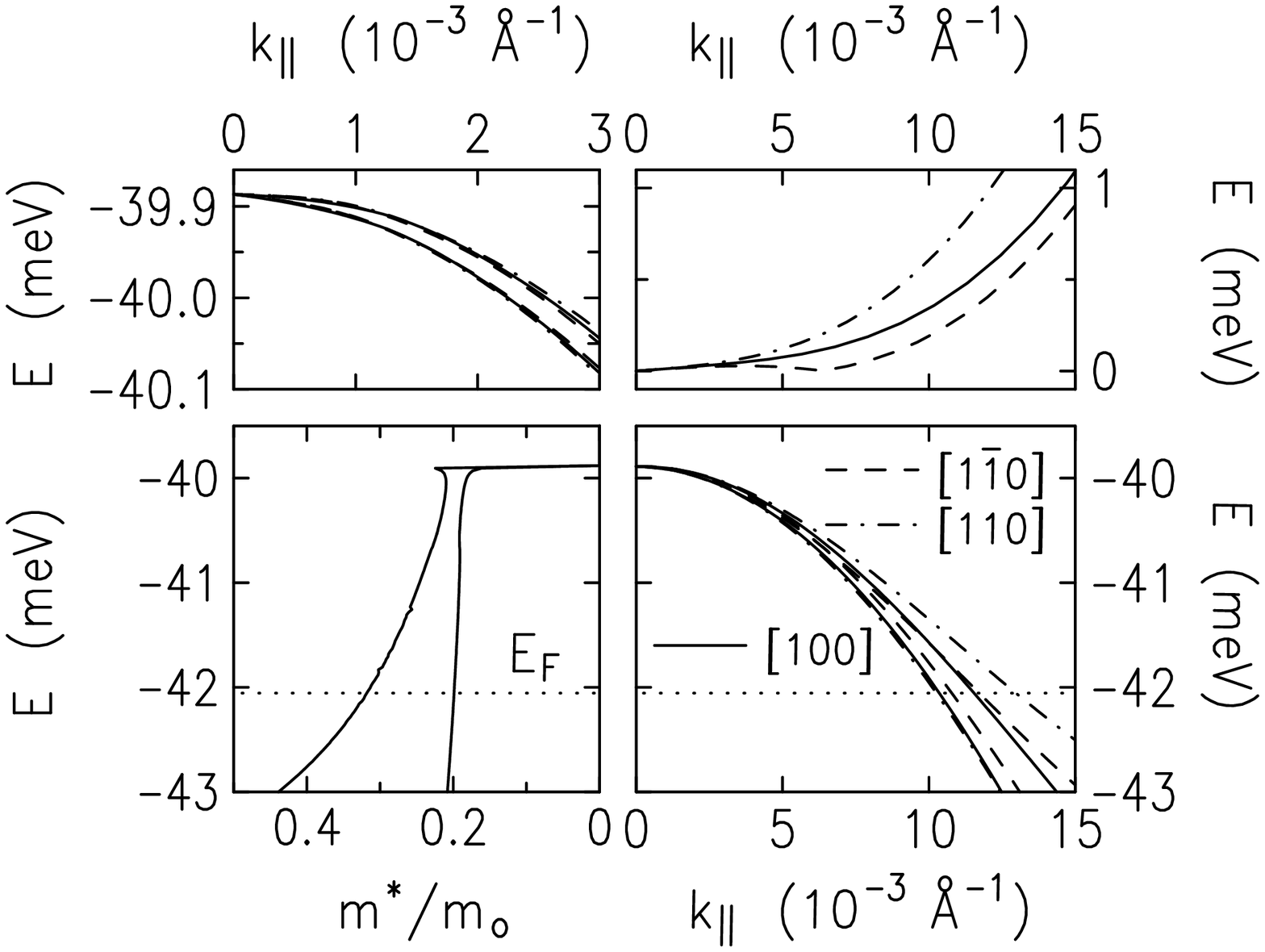}} \vspace{3mm}
\caption[]{\label{pic:gaas} Self-consistently calculated anisotropic
subband dispersion ${\cal E}_\pm ({\bf k}_\|)$ (lower right), DOS
effective mass $m^\ast/m_0$ (lower left), spin splitting ${\cal E}_+
({\bf k}_\|) - {\cal E}_- ({\bf k}_\|)$ (upper right) and subband
dispersion ${\cal E}_\pm ({\bf k}_\|)$ in the vicinity of $k_\|=0$
(upper left) for a [001] grown GaAs/Al$_{0.5}$Ga$_{0.5}$As
heterostructure with $N_s = 2 \times 10^{11}$ cm$^{-2}$ and $|N_A -
N_D| = 2\times 10^{16}$ cm$^{-2}$. Different line styles correspond
to different directions of the in-plane wave vector ${\bf k}_\|$ as
indicated. The dotted line indicates the Fermi energy $E_F$.}
\end{figure}

As an example, we show in Fig.\ \ref{pic:gaas} the self-consistently
calculated \cite{win93} anisotropic subband dispersion ${\cal E}_\pm
({\bf k}_\|)$, DOS effective mass $m^\ast/m_0$, and spin splitting
${\cal E}_+ ({\bf k}_\|) - {\cal E}_- ({\bf k}_\|)$ for a [001]
grown GaAs/Al$_{0.5}$Ga$_{0.5}$As heterostructure. The calculation
was based on a $14\times 14$ Hamiltonian ($\Gamma_8^c$,
$\Gamma_7^c$, $\Gamma_6^c$, $\Gamma_8^v$, and $\Gamma_7^v$).  It
fully took into account both SIA and BIA. The weakly divergent van
Hove singularity of the DOS effective mass at the subband edge
indicates that the $k$ linear splitting is rather small. (Its
dominant part is due to BIA. \cite{trr79}) Basically, the spin
splitting in Fig.\ \ref{pic:gaas} is proportional to $k_\|^3$.

Only for the crystallographic growth directions [001] and [111] the
hole subband states at $k_\| = 0$ are pure HH and LH states. For
low-symmetry growth directions like [113] and [110] we have mixed
HH-LH eigenstates even at $k_\| = 0$, though often the eigenstates
can be labeled by their dominant spinor components. \cite{win96a}
The HH-LH mixing adds a $k$ linear term to the splitting
(\ref{luttperturbHH}) of the HH states, which often exceeds $\langle
\beta_2 E_z \rangle k_\|$. However, this effect is still small when
compared with the cubic splitting.

For a HH subband dispersion ${\cal E}_\pm (k_\|) = \langle \mu
\rangle k_\|^2 \pm \langle \beta_1 E_z \rangle k_\|^3$ we obtain for
the densities $N_\pm$ in the spin-split subbands
\begin{mathletters}
\label{sdh_hh}
\begin{equation}
N_\pm = \frac{1}{2} N_s \pm \frac{\langle \beta_1 E_z \rangle 
N_s}{\sqrt{2} \, \langle \mu \rangle  X} \sqrt{\pi N_s (6 - 4/X)}
\end{equation}
with
\begin{equation}
X = 1 + \sqrt{1 - 4\pi \, N_s \left(\frac{\langle \beta_1
  E_z \rangle}{\langle \mu \rangle} \right)^2} .
\end{equation}
\end{mathletters}%
The spin splitting according to Eq.\ (\ref{sdh_hh}) is substantially
different from Eq.\ (\ref{sdh_rashba}). For electrons and a fixed
electric field $E_z$ but varying $N_s$ the difference $\Delta N =
N_+ - N_-$ increases like $N_s^{1/2}$ whereas for HH subbands it
increases like $N_s^{3/2}$. Using a fixed density $N_s$ but varying
$E_z$ it is more difficult to detect the difference between Eqs.\ 
(\ref{sdh_rashba}) and (\ref{sdh_hh}). In both cases a power
expansion of $\Delta N$ gives $\Delta N = a_1 |E_z| + a_3 |E_z|^3 +
{\cal O} (|E_z|^5)$ with $a_3 <0$ for electrons and $a_3>0$ for HH
subbands.

The proportionality (\ref{so_bulk}) is completely analogous to the
effective $g$ factor in bulk semiconductors. \cite{rot59} Lassnig
\cite{las85} pointed out that the $B=0$ spin splitting of electrons
can be expressed in terms of a position dependent effective $g$
factor $g^\ast (z)$. In the following we want to discuss the close
relationship between Zeeman splitting and $B=0$ spin splitting from
a more general point of view. Note that in the presence of an
external magnetic field ${\bf B}$ we have ${\bf k} \times {\bf k} =
(- i e / \hbar) {\bf B}$ and the Zeeman splitting in the
$\Gamma_6^c$ conduction band can be expressed as \cite{trr79}
\begin{equation}
\label{zeeman6c}
H_{6c}^{Z} = \frac{i\hbar}{e} \, \frac{g^\ast}{2} \, \mu_B \, {\bf
k} \times {\bf k} \cdot \bbox{\sigma}  =
\frac{g^\ast}{2} \, \mu_B \, {\bf B} \cdot \bbox{\sigma}
\end{equation}
with $\mu_B$ the Bohr magneton. Thus apart from a prefactor we
obtain the Rashba term (\ref{rashbafull}) from Eq.\ (\ref{zeeman6c})
by replacing one of the ${\bf k}$'s with the electric field ${\bf
E}$. In the $\Gamma_8^v$ valence band we have two invariants for the
Zeeman splitting \cite{lut56,trr79}
\begin{equation}
\label{lutt_zeeman}
H_{8v}^{Z} = 2 \kappa \, \mu_B \, {\bf B} \cdot {\bf J} +
2 q \, \mu_B \, {\bf B} \cdot \bbox{\cal J} .
\end{equation}
Here, the first term is the isotropic contribution, and the second
term is the anisotropic part. It is well-known that in all common
semiconductors for which Eq.\ (\ref{lutt_zeeman}) is applicable the
dominant contribution to $H_{8v}^{Z}$ is given by the first term
proportional to $\kappa$ whereas the second term is rather small.
\cite{trr79,lut56} Analogous to $\beta_1$ and $\beta_2$ the
isotropic ${\bf k} \cdot {\bf p}$ coupling between $\Gamma_8^v$ and
$\Gamma_6^c$ contributes to $\kappa$ but not to $q$. The latter
stems from ${\bf k} \cdot {\bf p}$ coupling to more remote bands
such as $\Gamma_8^c$ and $\Gamma_7^c$.

Several authors \cite{pud97,nit97,hei98,eis84} used an apparently
closely related intuitive picture for the $B=0$ spin splitting which
was based on the idea that the velocity $v_\| = \hbar k_\| / m^\ast$
of the 2D electrons is perpendicular to the electric field $E_z$. In
the electron's rest frame $E_z$ is Lorentz transformed into a
magnetic field $B$ so that the $B=0$ spin splitting becomes a Zeeman
splitting in the electron's rest frame. However, this magnetic field
is given by $B=(v_\|/c^2)E_z$ (SI units) and for typical values of
$E_z$ and $v_\|$ we have $B \sim 2 \ldots 20 \times 10^{-7}$~T which
would result in a spin splitting of the order of $5\times 10^{-9}
\ldots 5\times 10^{-5}$~meV. On the other hand, the experimentally
observed spin splitting is of the order of $0.1 \ldots 10$~meV. The
$B=0$ spin splitting requires the SO interaction caused by the
atomic cores. In bulk semiconductors this interaction is responsible
for the SO gap $\Delta_0$ between the valence bands $\Gamma_8^v$ and
$\Gamma_7^v$ which appears in Eq.\ (\ref{so_bulk}). The SO
interaction is the larger the larger the atomic number of the
constituting atoms. In Si we have $\Delta_0 = 44$ meV whereas in Ge
we have $\Delta_0 = 296$ meV. Therefore, SIA spin splitting in Si
quantum structures is rather small. \cite{win96}

Recently, spin splitting in 2D systems has gained renewed interest
because of an argument by Pudalov \cite{pud97} which relates the
metal-insulator transition (MIT) in low-density 2D systems with the
SIA spin splitting. Based on the Rashba model \cite{bych84} it was
argued that the SIA spin splitting ``results in a drastic change of
the internal properties of the system even without allowing for the
Coulomb interaction.'' \cite{sko98}. However, as we have shown
above, this argument is applicable only to electron and LH states.
The MIT has been observed also in pure HH systems in, e.g., Si/SiGe
QW's. \cite{lam97,col97} As noted above, SO interaction and spin
splitting in these systems are rather small, \cite{win96} so that it
appears unlikely that here the broken inversion symmetry of the
confining potential is responsible for the MIT. We note that in Si
2D electron systems the effective $g$ factor is enhanced due to many
body effects. \cite{afs82,oka99} It can be expected that similar
effects are also relevant for the $B=0$ spin splitting.

In conclusion, we have analyzed the SIA spin splitting in 2D
electron and hole systems. In 2D hole systems the splitting differs
remarkably from its familiar counterpart in the conduction band. For
electron states it increases linearly with in-plane wave vector
$k_\|$ whereas the spin splitting of heavy hole states can be of
third order in $k_\|$. We have discussed consequences of this
behavior in the context of recent arguments on the origin of the
metal-insulator transition observed in 2D systems.

The author wants to thank O.\ Pankratov, S.\ J.\ Papadakis, and M.\
Shayegan for stimulating discussions and suggestions.



\widetext

\begin{thebibliography}{99}
\bibitem{cross} In the same way we can derive $H^{\rm SO}$ for other
 blocks of a multiband Hamiltonian, \cite{trr79} e.g., for
 $H_{7v}^{\rm SO}$ we get Eq.\ (\ref{rashbafull}).
\bibitem{gap} For a third-order term such as $\langle
 h_i|k_\pm|c_j\rangle \cdot \langle c_j|k_z|h_k\rangle \cdot \langle
 h_k|E_z z|h_i\rangle$ we have only one energy denominator of the
 order of the fundamental gap $E_g$; the second denominator is given
 by the subband spacing. Note that $1/m^\ast \propto 1/E_g$, i.e.,
 both spin splitting and $\hbar^2 k_\|^2/(2m^\ast)$ scale like
 $E_g^{-1}$. If we restrict ourselves to the Luttinger Hamiltonian
 we get Eq.\ (\ref{lutt_spin_x}) from a second order perturbation
 theory for $E_z z$ and the off-diagonal terms $\gamma_3 \, k_\pm
 k_z$.  Here, the energy denominator $E_g$ is contained in the
 Luttinger parameter $\gamma_3$.
\end{thebibliography}
\end{document}